\documentclass[twocolumn,letterpaper,aps,prc,longbibliography,superscriptaddress,showpacs,nofootinbib,floatfix]{revtex4-1}

\usepackage{graphicx}	
\usepackage{xspace}	

\newcommand{\pt}{\mbox{$p_T$}\xspace}

\newcommand{\Ncoll}{\mbox{$N_{\rm coll}$}\xspace}

\newcommand{\sqs}{\mbox{$\sqrt{s}$}\xspace}
\newcommand{\sqsn}{\mbox{$\sqrt{s_{_{NN}}}$}\xspace}
\newcommand{\sqsntwo}{\mbox{$\sqrt{s_{_{NN}}}=200$~GeV}\xspace}
\newcommand{\pp}{\mbox{$p$$+$$p$}\xspace}

\newcommand{\pa}{\mbox{$p$$+$A}\xspace}
\newcommand{\pau}{\mbox{$p$$+$Au}\xspace}
\newcommand{\pal}{\mbox{$p$$+$Al}\xspace}

\newcommand{\pout}{\mbox{$p_{\rm out}$}\xspace}
\newcommand{\kt}{\mbox{$k_T$}\xspace}
\newcommand{\jt}{\mbox{$j_T$}\xspace}

\newcommand{\gevc}{\mbox{GeV/$c$}\xspace}
\newcommand{\xe}{\mbox{$x_E$}\xspace}
\newcommand{\dphi}{\mbox{$\Delta\phi$}\xspace}
\newcommand{\ptassoc}{\mbox{$p_T^{\rm assoc}$}\xspace}
\newcommand{\pttrig}{\mbox{$p_T^{\rm trig}$}\xspace}
\newcommand{\pion}{\mbox{$\pi^0$}\xspace}

\begin{document}

\title{Nonperturbative-transverse-momentum broadening in dihadron 
angular correlations in $\sqrt{s_{_{NN}}}=200$ GeV proton-nucleus 
collisions}

\newcommand{\abilene}{Abilene Christian University, Abilene, Texas 79699, USA}
\newcommand{\augie}{Department of Physics, Augustana University, Sioux Falls, South Dakota 57197, USA}
\newcommand{\banaras}{Department of Physics, Banaras Hindu University, Varanasi 221005, India}
\newcommand{\barc}{Bhabha Atomic Research Centre, Bombay 400 085, India}
\newcommand{\baruch}{Baruch College, City University of New York, New York, New York, 10010 USA}
\newcommand{\bnlcoll}{Collider-Accelerator Department, Brookhaven National Laboratory, Upton, New York 11973-5000, USA}
\newcommand{\bnlphys}{Physics Department, Brookhaven National Laboratory, Upton, New York 11973-5000, USA}
\newcommand{\caucr}{University of California-Riverside, Riverside, California 92521, USA}
\newcommand{\charlesczech}{Charles University, Ovocn\'{y} trh 5, Praha 1, 116 36, Prague, Czech Republic}
\newcommand{\chonbuk}{Chonbuk National University, Jeonju, 561-756, Korea}
\newcommand{\cns}{Center for Nuclear Study, Graduate School of Science, University of Tokyo, 7-3-1 Hongo, Bunkyo, Tokyo 113-0033, Japan}
\newcommand{\colorado}{University of Colorado, Boulder, Colorado 80309, USA}
\newcommand{\columbia}{Columbia University, New York, New York 10027 and Nevis Laboratories, Irvington, New York 10533, USA}
\newcommand{\czechtech}{Czech Technical University, Zikova 4, 166 36 Prague 6, Czech Republic}
\newcommand{\debrecen}{Debrecen University, H-4010 Debrecen, Egyetem t{\'e}r 1, Hungary}
\newcommand{\elte}{ELTE, E{\"o}tv{\"o}s Lor{\'a}nd University, H-1117 Budapest, P{\'a}zm{\'a}ny P.~s.~1/A, Hungary}
\newcommand{\eszterhazy}{Eszterh\'azy K\'aroly University, K\'aroly R\'obert Campus, H-3200 Gy\"ongy\"os, M\'atrai \'ut 36, Hungary}
\newcommand{\ewha}{Ewha Womans University, Seoul 120-750, Korea}
\newcommand{\fsu}{Florida State University, Tallahassee, Florida 32306, USA}
\newcommand{\gsu}{Georgia State University, Atlanta, Georgia 30303, USA}
\newcommand{\hiroshima}{Hiroshima University, Kagamiyama, Higashi-Hiroshima 739-8526, Japan}
\newcommand{\howard}{Department of Physics and Astronomy, Howard University, Washington, DC 20059, USA}
\newcommand{\ihepprot}{IHEP Protvino, State Research Center of Russian Federation, Institute for High Energy Physics, Protvino, 142281, Russia}
\newcommand{\illuiuc}{University of Illinois at Urbana-Champaign, Urbana, Illinois 61801, USA}
\newcommand{\inrras}{Institute for Nuclear Research of the Russian Academy of Sciences, prospekt 60-letiya Oktyabrya 7a, Moscow 117312, Russia}
\newcommand{\instpasczech}{Institute of Physics, Academy of Sciences of the Czech Republic, Na Slovance 2, 182 21 Prague 8, Czech Republic}
\newcommand{\isu}{Iowa State University, Ames, Iowa 50011, USA}
\newcommand{\jaea}{Advanced Science Research Center, Japan Atomic Energy Agency, 2-4 Shirakata Shirane, Tokai-mura, Naka-gun, Ibaraki-ken 319-1195, Japan}
\newcommand{\kek}{KEK, High Energy Accelerator Research Organization, Tsukuba, Ibaraki 305-0801, Japan}
\newcommand{\korea}{Korea University, Seoul, 02841, Korea}
\newcommand{\kurchatov}{National Research Center ``Kurchatov Institute", Moscow, 123098 Russia}
\newcommand{\kyoto}{Kyoto University, Kyoto 606-8502, Japan}
\newcommand{\lawllnl}{Lawrence Livermore National Laboratory, Livermore, California 94550, USA}
\newcommand{\losalamos}{Los Alamos National Laboratory, Los Alamos, New Mexico 87545, USA}
\newcommand{\lund}{Department of Physics, Lund University, Box 118, SE-221 00 Lund, Sweden}
\newcommand{\lyon}{IPNL, CNRS/IN2P3, Univ Lyon, Université Lyon 1, F-69622, Villeurbanne, France}
\newcommand{\maryland}{University of Maryland, College Park, Maryland 20742, USA}
\newcommand{\mass}{Department of Physics, University of Massachusetts, Amherst, Massachusetts 01003-9337, USA}
\newcommand{\michigan}{Department of Physics, University of Michigan, Ann Arbor, Michigan 48109-1040, USA}
\newcommand{\muhlenberg}{Muhlenberg College, Allentown, Pennsylvania 18104-5586, USA}
\newcommand{\nara}{Nara Women's University, Kita-uoya Nishi-machi Nara 630-8506, Japan}
\newcommand{\natmephi}{National Research Nuclear University, MEPhI, Moscow Engineering Physics Institute, Moscow, 115409, Russia}
\newcommand{\newmex}{University of New Mexico, Albuquerque, New Mexico 87131, USA}
\newcommand{\nmsu}{New Mexico State University, Las Cruces, New Mexico 88003, USA}
\newcommand{\ohio}{Department of Physics and Astronomy, Ohio University, Athens, Ohio 45701, USA}
\newcommand{\ornl}{Oak Ridge National Laboratory, Oak Ridge, Tennessee 37831, USA}
\newcommand{\orsay}{IPN-Orsay, Univ.~Paris-Sud, CNRS/IN2P3, Universit\'e Paris-Saclay, BP1, F-91406, Orsay, France}
\newcommand{\peking}{Peking University, Beijing 100871, People's Republic of China}
\newcommand{\pnpi}{PNPI, Petersburg Nuclear Physics Institute, Gatchina, Leningrad region, 188300, Russia}
\newcommand{\riken}{RIKEN Nishina Center for Accelerator-Based Science, Wako, Saitama 351-0198, Japan}
\newcommand{\rikjrbrc}{RIKEN BNL Research Center, Brookhaven National Laboratory, Upton, New York 11973-5000, USA}
\newcommand{\rikkyo}{Physics Department, Rikkyo University, 3-34-1 Nishi-Ikebukuro, Toshima, Tokyo 171-8501, Japan}
\newcommand{\saispbstu}{Saint Petersburg State Polytechnic University, St.~Petersburg, 195251 Russia}
\newcommand{\seoulnat}{Department of Physics and Astronomy, Seoul National University, Seoul 151-742, Korea}
\newcommand{\stonybrkc}{Chemistry Department, Stony Brook University, SUNY, Stony Brook, New York 11794-3400, USA}
\newcommand{\stonycrkp}{Department of Physics and Astronomy, Stony Brook University, SUNY, Stony Brook, New York 11794-3800, USA}
\newcommand{\tenn}{University of Tennessee, Knoxville, Tennessee 37996, USA}
\newcommand{\titech}{Department of Physics, Tokyo Institute of Technology, Oh-okayama, Meguro, Tokyo 152-8551, Japan}
\newcommand{\tsukuba}{Tomonaga Center for the History of the Universe, University of Tsukuba, Tsukuba, Ibaraki 305, Japan}
\newcommand{\vandy}{Vanderbilt University, Nashville, Tennessee 37235, USA}
\newcommand{\weizmann}{Weizmann Institute, Rehovot 76100, Israel}
\newcommand{\wigner}{Institute for Particle and Nuclear Physics, Wigner Research Centre for Physics, Hungarian Academy of Sciences (Wigner RCP, RMKI) H-1525 Budapest 114, POBox 49, Budapest, Hungary}
\newcommand{\yonsei}{Yonsei University, IPAP, Seoul 120-749, Korea}
\newcommand{\zagreb}{Department of Physics, Faculty of Science, University of Zagreb, Bijeni\v{c}ka c.~32 HR-10002 Zagreb, Croatia}
\affiliation{\abilene}
\affiliation{\augie}
\affiliation{\banaras}
\affiliation{\barc}
\affiliation{\baruch}
\affiliation{\bnlcoll}
\affiliation{\bnlphys}
\affiliation{\caucr}
\affiliation{\charlesczech}
\affiliation{\chonbuk}
\affiliation{\cns}
\affiliation{\colorado}
\affiliation{\columbia}
\affiliation{\czechtech}
\affiliation{\debrecen}
\affiliation{\elte}
\affiliation{\eszterhazy}
\affiliation{\ewha}
\affiliation{\fsu}
\affiliation{\gsu}
\affiliation{\hiroshima}
\affiliation{\howard}
\affiliation{\ihepprot}
\affiliation{\illuiuc}
\affiliation{\inrras}
\affiliation{\instpasczech}
\affiliation{\isu}
\affiliation{\jaea}
\affiliation{\kek}
\affiliation{\korea}
\affiliation{\kurchatov}
\affiliation{\kyoto}
\affiliation{\lawllnl}
\affiliation{\losalamos}
\affiliation{\lund}
\affiliation{\lyon}
\affiliation{\maryland}
\affiliation{\mass}
\affiliation{\michigan}
\affiliation{\muhlenberg}
\affiliation{\nara}
\affiliation{\natmephi}
\affiliation{\newmex}
\affiliation{\nmsu}
\affiliation{\ohio}
\affiliation{\ornl}
\affiliation{\orsay}
\affiliation{\peking}
\affiliation{\pnpi}
\affiliation{\riken}
\affiliation{\rikjrbrc}
\affiliation{\rikkyo}
\affiliation{\saispbstu}
\affiliation{\seoulnat}
\affiliation{\stonybrkc}
\affiliation{\stonycrkp}
\affiliation{\tenn}
\affiliation{\titech}
\affiliation{\tsukuba}
\affiliation{\vandy}
\affiliation{\weizmann}
\affiliation{\wigner}
\affiliation{\yonsei}
\affiliation{\zagreb}
\author{C.~Aidala} \affiliation{\michigan} 
\author{Y.~Akiba} \email[PHENIX Spokesperson: ]{akiba@rcf.rhic.bnl.gov} \affiliation{\riken} \affiliation{\rikjrbrc} 
\author{M.~Alfred} \affiliation{\howard} 
\author{V.~Andrieux} \affiliation{\michigan} 
\author{N.~Apadula} \affiliation{\isu} 
\author{H.~Asano} \affiliation{\kyoto} \affiliation{\riken} 
\author{B.~Azmoun} \affiliation{\bnlphys} 
\author{V.~Babintsev} \affiliation{\ihepprot} 
\author{N.S.~Bandara} \affiliation{\mass} 
\author{K.N.~Barish} \affiliation{\caucr} 
% ...
\collaboration{PHENIX Collaboration}  \noaffiliation

\date{\today}


\begin{abstract}

The PHENIX collaboration has measured high-$p_T$ dihadron correlations 
in $p$$+$$p$, $p$$+$Al, and $p$$+$Au collisions at 
$\sqrt{s_{_{NN}}}=200$ GeV. The correlations arise from inter- and 
intra-jet correlations and thus have sensitivity to nonperturbative 
effects in both the initial and final states. The distributions of 
$p_{\rm out}$, the transverse momentum component of the associated 
hadron perpendicular to the trigger hadron, are sensitive to initial and 
final state transverse momenta. These distributions are measured 
multi-differentially as a function of $x_E$, the longitudinal momentum 
fraction of the associated hadron with respect to the trigger hadron. 
The near-side $p_{\rm out}$ widths, sensitive to fragmentation 
transverse momentum, show no significant broadening between $p$$+$Au, 
$p$$+$Al, and $p$$+$$p$. The away-side nonperturbative $p_{\rm out}$ 
widths are found to be broadened in $p$$+$Au when compared to $p$$+$$p$; 
however, there is no significant broadening in $p$$+$Al compared to 
$p$$+$$p$ collisions. The data also suggest that the away-side $p_{\rm 
out}$ broadening is a function of $N_{\rm coll}$, the number of binary 
nucleon-nucleon collisions, in the interaction. The potential 
implications of these results with regard to initial and final state 
transverse momentum broadening and energy loss of partons in a nucleus, 
among other nuclear effects, are discussed.

\end{abstract}

\maketitle
          
\section{Introduction}

High energy collisions of protons with nuclei provide a testing ground 
for quantum chromodynamics (QCD). In particular, when large transverse 
momentum scales are involved, the collisions can probe the quark and 
gluon, collectively referred to as partons, structure of the nucleus. 
Proton-nucleus (\pa) collisions have traditionally been used as a 
control to identify final-state nuclear effects in high energy 
nucleus-nucleus collisions where a strongly interacting quark-gluon 
plasma (QGP) is formed~\cite{Adcox:2004mh}. However, measurements in \pa 
collisions have revealed many surprising results that have yet to be 
completely reconciled with each other; these have shown that 
understanding and explaining many different ``cold'' nuclear matter 
effects is already a challenging 
endeavor~\cite{Hen:2016kwk,CMS:2012qk,Cronin:1974zm,Adcox:2001jp}.

For example, in the initial-state, nuclei are known to be more complex 
than just a simple linear superposition of nucleons (see e.g. 
Ref.~\cite{Hen:2016kwk} for a review). Nuclear parton distribution 
functions (PDFs) are known to have several regions where they deviate 
from simple superpositions of nucleon PDFs as a function of the 
longitudinal momentum fraction $x$ that the parton carries of the 
nucleon. Understanding how the partonic degrees of freedom lead to 
nuclear structure will be a major achievement of QCD; however, there is 
still significant effort required in understanding the physical origin 
of these measured nuclear modifications. Final-state hadronization from 
a nucleus can also be modified similarly to nuclear PDFs in the initial 
state. In particular, semi-inclusive deep-inelastic scattering (SIDIS) 
experiments have shown that high $z$ hadrons are suppressed in 
electron-nucleus relative to electron-deuterium 
collisions~\cite{Airapetian:2003mi}, where $z$ is the longitudinal 
momentum fraction of the outgoing hadron with respect to the fragmenting 
parton. This suppression was found to be dependent on the nuclear target 
size~\cite{Airapetian:2007vu}. In addition, a particle species 
dependence was observed, which may reflect differences in the nuclear 
modification of quark and/or antiquark fragmentation functions and 
possible differences in meson versus baryon production from 
nuclei~\cite{Airapetian:2003mi,Airapetian:2007vu}.

Several proposed signatures of the QGP have also been measured in \pa 
collisions where the overall system size created in the collision was 
once expected to be too small. Collective behavior has been observed 
across large pseudorapidity ranges in high multiplicity \pa 
collisions~\cite{CMS:2012qk,Aad:2012gla,Aaij:2015qcq,Adamczyk:2015xjc,Aidala:2016vgl,Aidala:2017ajz}. 
Additionally, the enhancement of strangeness in hadron production in 
high multiplicity \pa collisions has recently been 
measured~\cite{Adam:2015vsf}. Surprisingly, both of these phenomena have 
also been observed in high multiplicity \pp collisions at 
Large-Hadron-Collider (LHC) energies~\cite{Aad:2015gqa,ALICE:2017jyt}. 
However, the suppression of high \pt inclusive hadrons or jets in \pa 
collisions with respect to \pp collisions has not been 
measured~\cite{Adler:2003ii,ATLAS:2014cpa}. These results were first 
used to establish final-state QGP interactions as the cause of high \pt 
hadron suppression in nucleus-nucleus 
collisions~\cite{Adcox:2001jp,Chatrchyan:2011sx}. However, the recent 
addition of collective and strange hadron measurements but lack of 
hadron suppression in \pa collisions has complicated the idea that a QGP 
may be formed in smaller collision systems.

Another unexpected physical effect in \pa collisions is the so-called 
``Cronin'' effect, which refers to an enhancement in the inclusive 
hadron \pt spectrum with respect to \pp collisions at moderate \pt of 
approximately $2<\pt<6$ \gevc which persists over a wide range of 
center-of-mass energies~\cite{Cronin:1974zm,Adler:2003ii,Aad:2016zif}. 
This effect has also been observed at moderate \pt in electron-nucleus 
collisions~\cite{Airapetian:2003mi}, where a significant dependence of 
the enhancement on the longitudinal momentum fraction $z$ of the hadron 
was found~\cite{Airapetian:2007vu}. While this was first proposed to be 
due to multiple scattering effects in the nuclear medium, more recent 
measurements have shown that hadronization also plays a 
role~\cite{Adare:2013esx}. Additional measurements that go beyond single 
inclusive hadrons may be able to shed further light on this phenomenon 
in \pa collisions. For example, dijet measurements in the kinematic 
regime of the Cronin effect have shown that the initial-state partonic 
transverse momentum is a function of the nucleus 
size~\cite{Corcoran:1990vq}, which has not been observed at large jet 
transverse momentum~\cite{Adam:2015xea}.

The lack of understanding of the underlying physical sources of these 
phenomena motivates measurements in new kinematic regimes with different 
observables. Here we present a measurement of dihadron angular 
correlations in \pp, \pal, and \pau collisions at midrapidity collected 
by the PHENIX experiment at the Relativistic Heavy Ion Collider (RHIC). 
The unique capabilities of RHIC allow for a nuclear size dependence to 
be studied in \sqsn=~200 GeV \pa collisions. High \pt two-particle 
angular correlations have been theoretically considered as a probe for 
energy loss in \pa and A+A collisions via their transverse momentum 
broadening~\cite{Baier:1996sk,Baier:2000mf,Tannenbaum:2017afg}; however, 
the various aforementioned effects should also be considered as they 
will play a role in both collision systems. The present measurements 
will contribute to our understanding of the rich phenomena in hadronic 
interactions involving nuclei.

\section{Methods}

In 2015, the PHENIX experiment~\cite{Adcox:2003zm} at RHIC collected 
data from \pp, \pal, and \pau collisions at \sqsntwo. A total minimum 
bias integrated luminosity of 60, 0.69, and 0.21 pb$^{-1}$ for \pp, 
\pal, and \pau, respectively, was used for the analysis of dihadron 
correlations. From these total integrated luminosities, data quality 
assurance and collision vertex position $|z_{\rm vtx}|<30$ cm cuts were 
applied. The PHENIX detector measures two-particle angular correlations 
of neutral pions and charged hadrons, \pion-h$^{\pm}$, with its 
electromagnetic calorimeter (EMCal) and drift chamber (DC) and pad 
chamber (PC) tracking system. These central arms cover an azimuthal 
range of \dphi~$\approx\pi$ radians and a pseudorapidity 
range of $|\eta|<0.35$.  Detailed descriptions of the PHENIX central 
arms can be found in Refs.~\cite{Adcox:2003zp,Aphecetche:2003zr}. In \pa 
collisions, the centrality class is determined with the forward 
beam-beam counters (BBCs)~\cite{Ikematsu:1998fm}, where the centrality 
percentiles are defined by the multiplicity measured in the 
nucleus-going BBC following the procedure in Ref.~\cite{Adare:2013nff}.

The EMCal is used to identify high \pt neutral pions to construct the 
correlation functions. A high-energy-photon trigger is used to identify 
events with a high \pt photon from a $\pion\rightarrow\gamma\gamma$ 
decay. Photons are identified using a shower shape cut that removes 
charged hadrons as well as most high energy clusters that overlap with 
another photon. The neutral pions are reconstructed via their two photon 
decay, where the granularity of the EMCal allows \pion reconstruction up 
to approximately 20 GeV in this channel. In this analysis neutral pions 
are collected in the \pt range $5<\pt<9$ \gevc. 

The DC and PC tracking system measures nonidentified charged hadrons in 
the event with the triggered high \pt photon. Two PCs, located radially 
behind the DC, are used to identify and match tracks in the DC with hits 
in the PCs. This track matching condition reduces background from 
secondary tracks due to conversions or decays. A ring-imaging 
$\check{\textrm{C}}$erenkov system is also used to reject electrons from 
the charged hadron sample. With these conditions, the DC and PC tracking 
system is also used to reject tracks in the EMCal that happen to shower 
and are thus background for the $\pion\rightarrow\gamma\gamma$ 
identification. Nonidentified charged hadrons are collected between 
$0.5<\pt<10$ \gevc in correlation with the high \pt \pion to 
study the correlations in as large a range as allowed by the data.

The correlations are constructed similarly to previous PHENIX 
two-particle correlation analyses, see e.g. 
Refs.~\cite{Adler:2006sc,Adare:2010yw,Aidala:2018bjf}. Per-trigger yields 
are constructed for a given observable, such as \dphi, which show the yield 
of charged hadrons per-trigger \pion and are defined by

\begin{equation}\label{eq:pty}
\frac{1}{N_{\rm trig}}\frac{dN}{d\dphi} = \frac{1}{N_{\rm trig}}\frac{dN/d\dphi_{\rm raw}}{dN/d\dphi_{\rm mixed}\epsilon(\pt)}
\end{equation}

\noindent To account for the PHENIX acceptance, the raw correlations are 
divided by a mixed-event background correlation function, 
$dN/d\dphi_{\rm mixed}$. The background correlation is constructed with 
neutral pions and charged hadrons from the same data taking period but 
different event number; the events are required to have a similar 
centrality and z-vertex. To account for the efficiency of the PHENIX 
detector, the correlation functions are also corrected by a charged 
hadron efficiency defined as $\epsilon(\pt)$ in Eq.~\ref{eq:pty}, which 
is determined with a single particle Monte Carlo generator as well as a 
full GEANT description of the PHENIX apparatus~\cite{Adler:2003zv}. 
After these corrections, the correlations are normalized by the total 
number of trigger particles measured to construct the per-trigger yield 
and correspond to full azimuthal acceptance within $|\eta|<0.35$.

\begin{figure}[thb] 
	\includegraphics[width=1.0\linewidth]{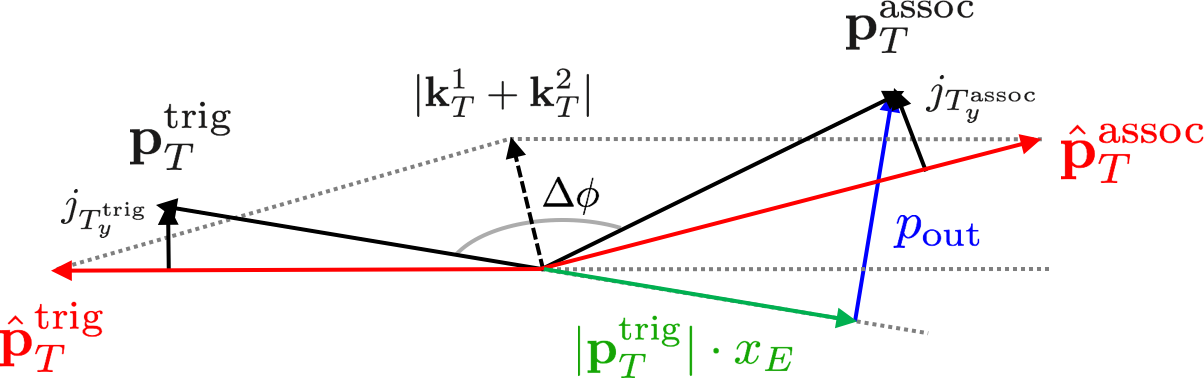}
\caption{A schematic diagram showing a dihadron correlation in 
the transverse plane. Vector quantities are shown in bold. The red 
vectors are the two partons, which are acoplanar due to initial-state 
partonic $k_T$, while the two black vectors are the measured trigger and 
associated hadron, slightly displaced from the partons due to 
final-state transverse momentum $j_T$ from fragmentation. The quantities \pout and \xe are 
shown as blue and green vectors, respectively.}
      \label{fig:ktkinematics}
\end{figure}              
              
Correlation functions are typically constructed in terms of the 
azimuthal angle \dphi between the trigger and associated particle. Here 
we choose to construct the correlations as a function of the momentum 
space vector component \pout and \xe, defined as

\begin{equation}
\pout = |{\bf{p}}_T^{\rm assoc}|\sin\dphi
\end{equation}
\noindent and
\begin{equation}
\xe = -\frac{{\bf{p}}_T^{\rm trig}\cdot{\bf{p}}_T^{\rm assoc}}{|{\bf{p}}_T^{\rm trig}|^2}=-\frac{|{\bf{p}}_T^{\rm assoc}|}{|{\bf{p}}_T^{\rm trig}|}\cos\dphi .
\end{equation}
              
\noindent The quantities \pout and \xe give the transverse momentum 
component and longitudinal momentum fraction, respectively, of the 
associated hadron with respect to the trigger \pion. These quantities 
are schematically diagrammed in Fig.~\ref{fig:ktkinematics}, where the 
Figure shows a two-particle correlation in the transverse plane and 
quantities in bold represent vectors. In this diagram, two hadrons 
(black vectors) fragment from two high \pt partons (red vectors) from a 
two-to-two partonic scattering. The partons are originally acoplanar due 
to their initial-state transverse momenta ($k_T$); the two hadrons may 
acquire additional acoplanarity due to final-state transverse momentum 
($j_T$) during the fragmentation process. In the diagram the final-state 
transverse momentum is perpendicular to the parton axis and denoted as 
$j_{T_{y}^{\rm trig}}$ and $j_{T_{y}^{\rm assoc}}$, which are assumed to 
be Gaussian such that $\sqrt{\langle j_T^2\rangle}=\sqrt{2\langle 
j^2_{T_{y}^{\rm trig}}\rangle}=\sqrt{2\langle j^2_{T_{y}^{\rm 
assoc}}\rangle}$. The quantity \jt could have a \pttrig or \ptassoc 
dependence to it; however, measurements have shown that this dependence 
is small~\cite{Acharya:2018edi}. The quantity \pout can be nonzero 
because of the $k_T$ and $j_T$ transverse momentum contributions, while 
\xe is a proxy for the momentum fraction $z$ that the final-state hadron 
carries with respect to the parton; see Fig.~\ref{fig:xevsz} and the 
associated text. In the figure, \xe is shown multiplied by \pttrig to 
explicitly show the comparison between \xe and $z$. When \pout is small, 
the two-particle correlation is nearly back-to-back and the acoplanarity 
is generated by nonperturbative \kt and \jt~\cite{Aidala:2018bjf,Adare:2016bug}. 
Additional nonperturbative interactions within the nucleus, as discussed 
and referenced in the Introduction, may contribute to this quantity in 
\pa collisions.
              
Systematic uncertainties are assigned for the charged hadron efficiency 
and for the underlying event background subtraction procedure. The 
systematic uncertainty on the charged hadron yields is determined to be 
an overall normalization uncertainty of 9\% on the per-trigger yields. 
The dominant contribution is due to the uncertainty that arises from 
matching tracks in the PHENIX drift chamber to the outermost pad 
chamber; however, there are also contributions from the overall tracking 
resolution of the detector and the Monte Carlo determination of the 
nonidentified charged hadron efficiency. The underlying event background 
is statistically subtracted with fits to the away-side \dphi correlation 
functions as described in Ref.~\cite{Adare:2016bug}; these fits 
determine the percentage of underlying event background level with 
respect to the jet yield. The systematic uncertainty is determined by 
altering the underlying event region by $\pm 1\sigma$ based on the fit 
results. This uncertainty varies from less than 1\% at small \pout to 
several percent at large $\pout$ where the underlying event 
contribution, and thus background-to-signal, is larger.

\section{Results}

\begin{figure}[thb]
	\includegraphics[width=1.0\linewidth]{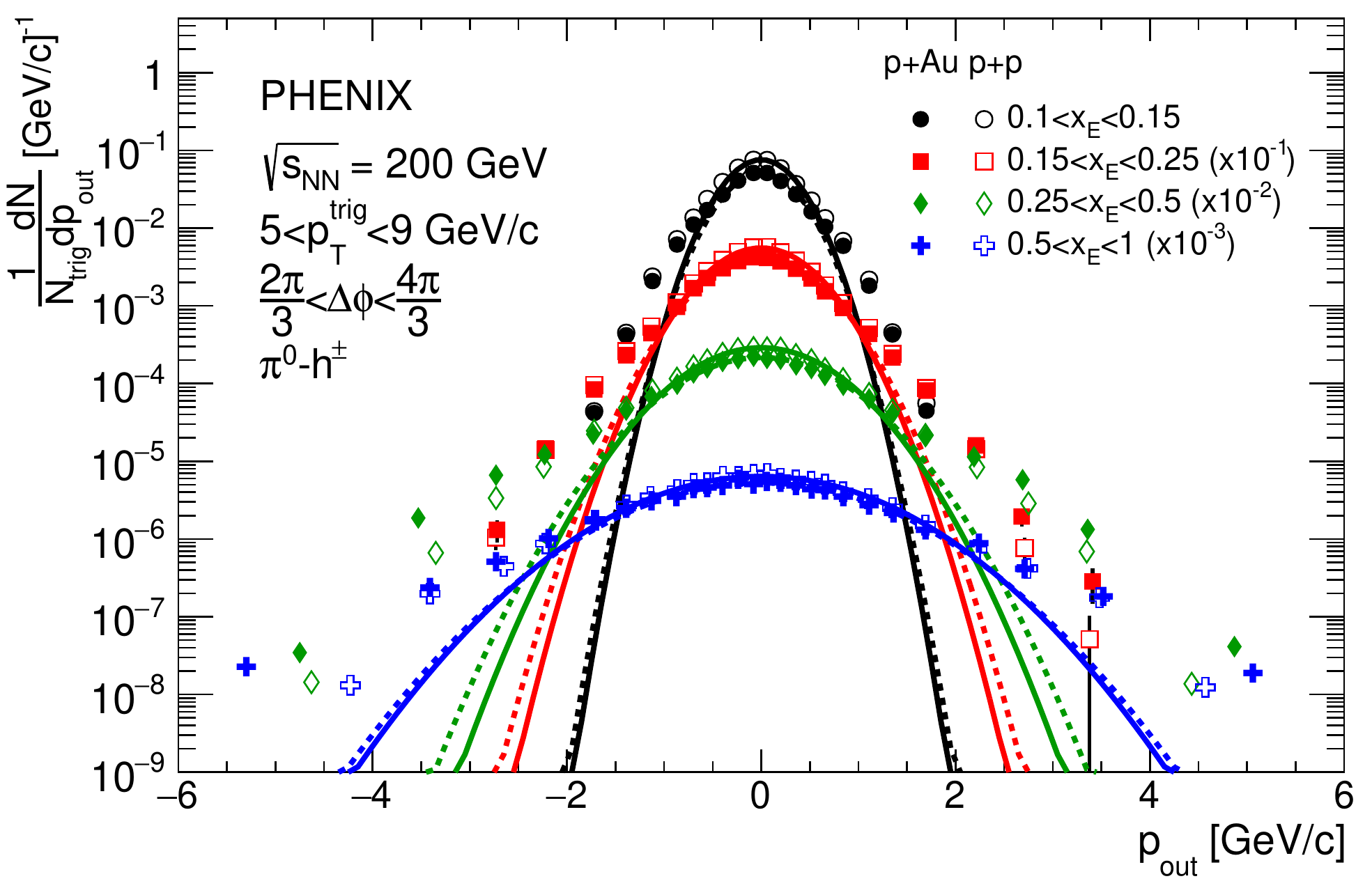}
\caption{The away-side \pout per-trigger yields are shown in both \pau 
and \pp collisions for several bins of \xe. Gaussian fits, shown as 
dotted lines for \pau and solid lines for \pp, are shown to the small 
\pout distributions, highlighting the nonperturbative to perturbative 
transition.}
     \label{fig:poutptys}
\end{figure}

Examples of the away-side per-trigger yields as a function of \pout are 
shown in several bins of \xe in Fig.~\ref{fig:poutptys}. The per-trigger 
yields for \pau and \pp collisions are shown as open and filled points, 
respectively. A transition from nonperturbative behavior in the nearly 
back-to-back \pout$\approx$0 (\dphi$\approx\pi$) region to perturbative 
next-to-leading order behavior at larger \pout can be seen at varying 
values of \pout, depending on the \xe bin. This change in shape is 
highlighted by Gaussian fits to the small \pout region, drawn in 
Fig.~\ref{fig:poutptys} as dotted and solid lines for \pau and \pp, 
respectively. The fit ranges vary depending on the \xe bin as the 
nonperturbative region is a function of \ptassoc and thus 
\xe~\cite{Aidala:2018bjf}. The fit ranges are chosen to give the best 
$\chi^2$/NDF, and a systematic uncertainty is assigned based on this 
choice as described later; this is the dominant systematic uncertainty 
on the Gaussian widths. Generally the fits have a $\chi^2$ per NDF of 
order $\sim$10 on the \pout distributions when fitting only the 
statistical uncertainties. When fitting the systematic uncertainties, 
the fits always have a $\chi^2$ per NDF less than unity. The fit results 
are nearly identical when fitting the statistical or systematic 
uncertainties; therefore, systematic uncertainties on the Gaussian 
widths are more conservatively estimated by adjusting the fit range when 
fitting to the statistical uncertainties. The fit ranges are the same 
for both the \pp and \pa data, and range from $[-0.4,0.4]$ GeV/$c$ in 
the smallest \xe bin to $[-1.5,1.5]$ GeV/$c$ in the largest \xe bin. 
These fits clearly do not describe the data at larger values of \pout 
where the data exhibit a more power-law like behavior.  This transition 
indicates a change from sensitivity to nonperturbative to perturbative 
physics effects.

\begin{figure}[thb]
    \includegraphics[width=1.0\linewidth]{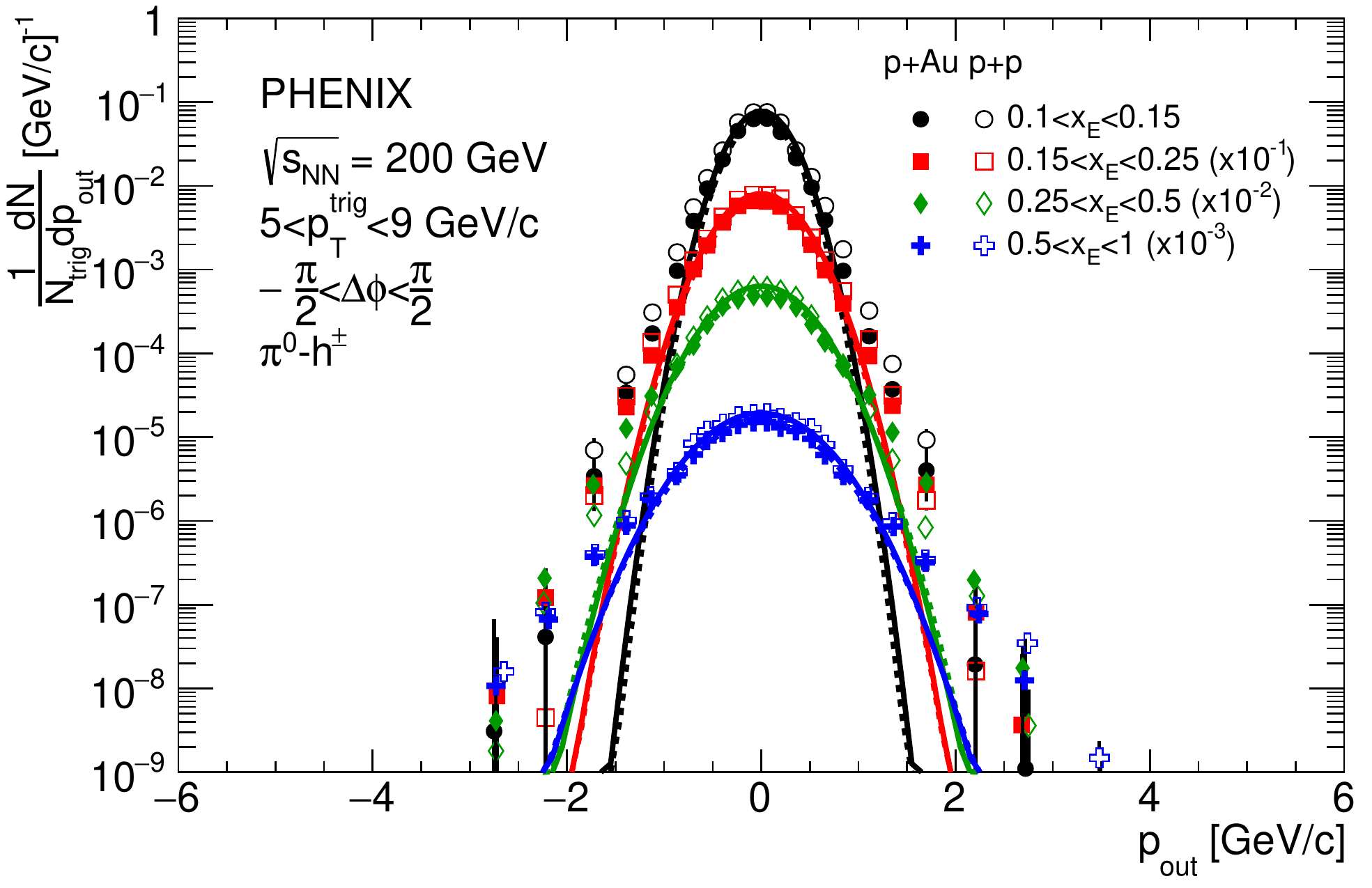}
\caption{The near-side \pout per-trigger yields are shown in both \pau 
and \pp collisions for several bins of \xe. Gaussian fits, shown as 
dotted lines for \pau and solid lines for \pp, are performed at small 
\pout where nonperturbative behavior is dominant in the $\dphi\approx$0 
region.}
    \label{fig:nearsidepouts}
\end{figure}

Examples of the near-side per-trigger yields as a function of \pout are 
shown in several bins of \xe in Fig.~\ref{fig:nearsidepouts}. The 
near-side per-trigger yields show a much narrower distribution than the 
away-side per-trigger yields due to the differences between intra-jet 
correlations and inter-jet correlations, respectively. In particular, 
the near-side correlations are only sensitive to fragmentation 
transverse momentum, because the \pion and hadron are fragmented from 
the same hard parton.  However, the away-side correlations are sensitive 
to both initial and final state transverse momentum.  Because the 
initial-state \kt is much larger than final-state \jt (see 
e.g.~\cite{Adler:2006sc,Adare:2010yw}), this leads to a broader \pout 
distribution on the away-side than the near-side. Nonetheless, a 
nonperturbative Gaussian region can still be identified on the near side 
as shown in Fig.~\ref{fig:nearsidepouts}, similarly to the away side, 
with a power law spectrum at larger \pout that is not well described by 
the Gaussian fit.

\begin{figure}[thb]
	\includegraphics[width=1.15\linewidth]{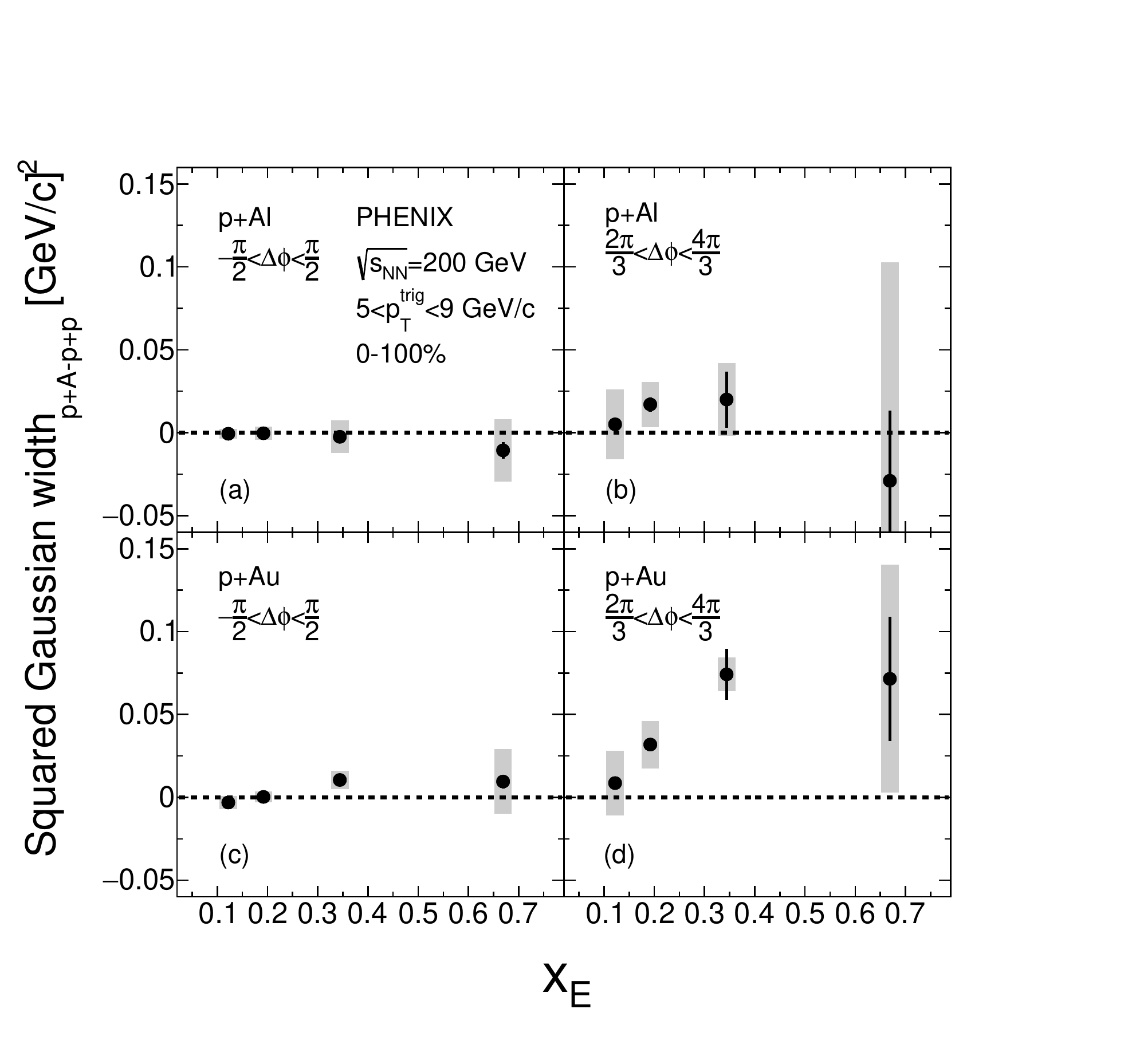}
\caption{The Gaussian width differences are shown for the near-side (a) 
and away-side (b) between \pal and \pp collisions and for the near-side 
(c) and away-side (d) between \pau and \pp collisions as a function of 
\xe.}
      \label{fig:quaddiff}
\end{figure}

To measure the nonperturbative momentum widths, the Gaussian widths are 
extracted from the fits to both the near and away side \pout 
distributions. Systematic uncertainties on the Gaussian widths are 
estimated by increasing the fit range by 0.2 \gevc in \pout and taking 
the absolute value of the difference of the resulting Gaussian width. To 
study any modification in \pa compared to \pp collisions the squared 
width difference is determined between the \pa and \pp Gaussian widths. 
These differences are shown in Fig.~\ref{fig:quaddiff} as a function of 
\xe for the near and away side correlation functions, for both \pal and 
\pau collisions. The near-side width differences in the left column of 
Fig.~\ref{fig:quaddiff} show no significant modification within 
uncertainties between both \pal or \pau and \pp collisions at all values 
of \xe. Similar results have been seen in dihadron 
correlations~\cite{Acharya:2018edi} and fragmentation function studies 
with full jet reconstruction~\cite{Aaboud:2017tke} in $p$$+$Pb 
collisions. However, the away-side width differences in \pau and \pp 
collisions show modification as seen in the bottom-right panel of 
Fig.~\ref{fig:quaddiff}. There is no significant away-side broadening 
between \pal and \pp collisions as seen in the top right panel of 
Fig.~\ref{fig:quaddiff} within the assigned systematic uncertainties.
      
\begin{figure}[thb]
	\includegraphics[width=1.0\linewidth]{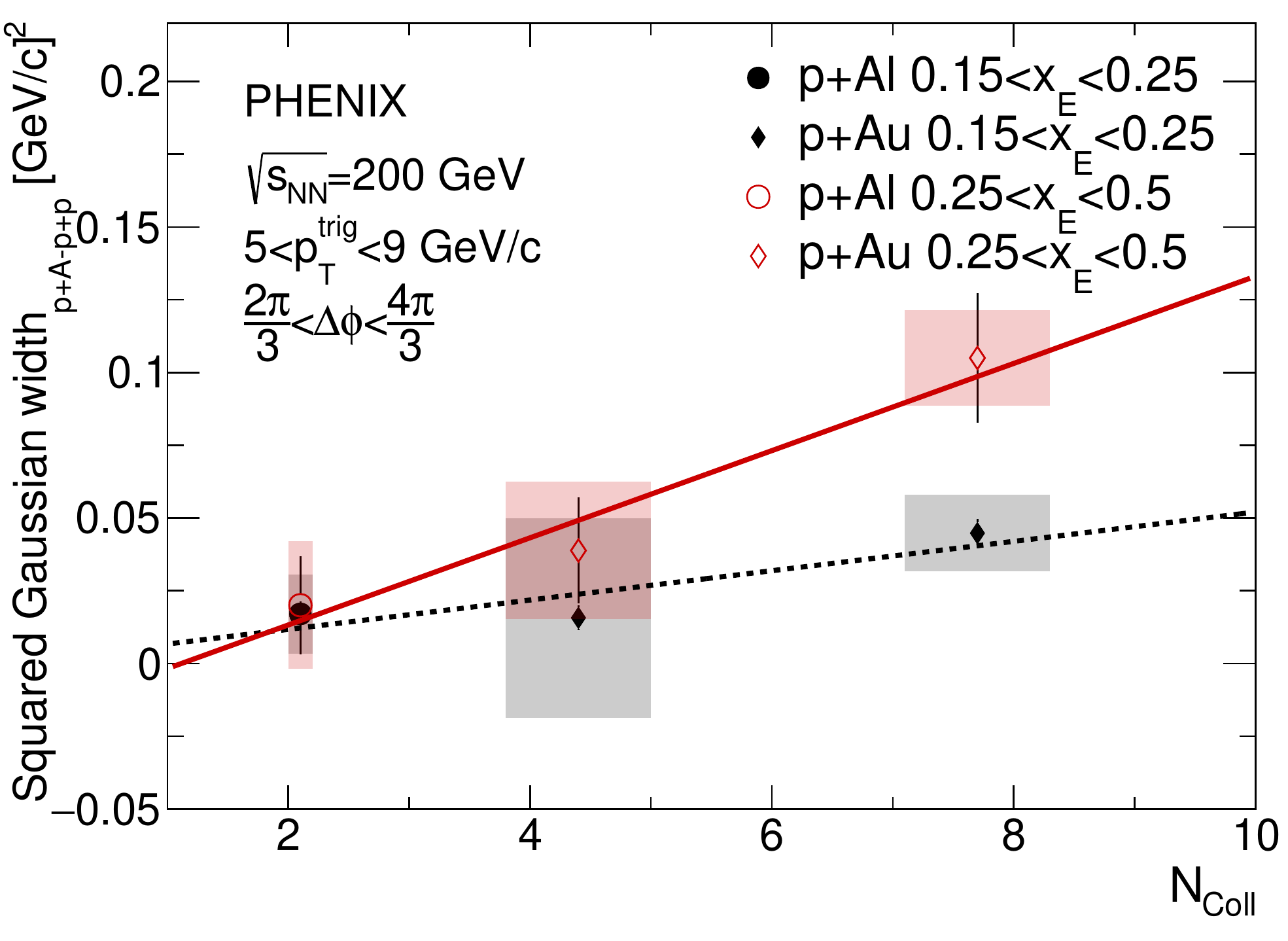}
\caption{The Gaussian width differences between \pa and \pp are shown in 
two \xe bins as a function of \Ncoll. Linear fits are shown for each \xe 
bin, which exhibit a positive dependence with \Ncoll.}
      \label{fig:widthsncoll}
\end{figure}

There is an indication that the away-side squared Gaussian width 
differences depend on the nucleus size as indicated in the right column 
of Fig.~\ref{fig:quaddiff}. To study this further, the \pout per-trigger 
yields were split into two centrality bins in the \pau data and the same 
analysis was performed. The centrality in \pa collisions is converted to 
values of \Ncoll with the method in Ref.~\cite{Adare:2013nff}, where 
\Ncoll is defined as the average number of nucleon-nucleon collisions in 
a given event class. Figure~\ref{fig:widthsncoll} shows the squared 
width differences in \pa and \pp collisions as a function of \Ncoll in 
the two \xe bins where a nonzero Gaussian width difference is observed. 
The values of these squared width differences are shown in 
Tab.~\ref{tab:pouts}. The data are fit with linear functions which are 
shown on the figure and indicate that the squared width differences 
exhibit a positive correlation with \Ncoll. The slopes of the fits were 
found to be $0.005\pm~0.001$~(stat)~$\pm0.003$~(sys) and 
0.015~$\pm~0.005$~(stat)~$\pm~0.004$~(sys) for the smaller and larger 
\xe bins, respectively. When the data is fit to a constant of 0, the 
$\chi^2$ per number of degree of freedom becomes approximately 5 for 
$0.15<\xe<0.25$ and approximately 8 for $0.25<\xe<0.5$. The measured 
slopes differ from a slope of 0 with p values of approximately 0.055 and 
0.01, for the smaller and larger \xe bin respectively, where the 
statistical and systematic uncertainties on the slopes were added in 
quadrature. This suggests that the interpretation of no transverse 
momentum broadening in \pa compared to \pp is inconsistent with the 
data.

\begin{table*}[tbh]
 \caption{\label{tab:pouts}
The values of the Gaussian width differences between \pa and \pp and 
their statistical and systematic uncertainties are shown, corresponding 
to Fig.~\ref{fig:widthsncoll}. Units are $[\rm{GeV}/c]^2$ for the width 
differences and their uncertainties.
}
 \begin{ruledtabular} \begin{tabular}{ccccccc}
  System &   \Ncoll & $\sigma_{\Ncoll}$ & $\xe$ &  Squared Gaussian width$_{~p+A-p+p}$ & (stat) & (syst) \\\hline
 \pal    & 2.1 & 0.1 & 0.15--0.25    &   0.017 &  0.004 &   0.013   \\
   \pau  & 4.4 & 0.6 & 0.15--0.25    &   0.016 &  0.004 &   0.034   \\
   \pau  & 7.7 & 0.6 & 0.15--0.25    &   0.045 &  0.005 &   0.013   \\
  \pal   & 2.1 & 0.1 & 0.25--0.50    &   0.020 &  0.017 &   0.022   \\
  \pau   & 4.4 & 0.6 & 0.25--0.50    &   0.039 &  0.018 &   0.023   \\
  \pau   & 7.7 & 0.6 & 0.25--0.50    &   0.105 &  0.022 &   0.016   \\
 \end{tabular} \end{ruledtabular}
 \end{table*}

\section{Discussion}

There are a number of different physical processes that could be 
contributing to the apparent broadening of the away-side nonperturbative 
momentum widths in \pa compared to \pp collisions, as discussed in the 
Introduction. The apparent lack of broadening on the near-side indicates 
that additional nonperturbative radiation during fragmentation in \pa is 
small. This may suggest that the fragmentation of the hard scattered 
parton into hadrons occurs outside any nuclear medium that is present; 
therefore, this fragmentation is similar between \pa and \pp collisions 
and is independent of the presence of a nucleus in the kinematic region 
probed by this data.

In the last decade, significant emphasis has been placed on the 
observation of collective effects in \pa collisions. The effects of 
contributions from $v_2$ and $v_3$ Fourier harmonics were studied and 
found to be negligible in the present analysis; this is because the 
Gaussian widths are almost entirely constrained by correlations in the 
range of $\dphi\pm0.2$ radians around $\dphi=\pi$. 
In this small $\dphi$ range, any modulation 
from collective dynamics was found to contribute on average less than 
1\% to the normalization of the \pout correlation functions. 
Additionally, the correlations are collected in the midrapidity 
$|\eta|<0.35$ region where the $\Delta\eta$ between the high \pt trigger 
and associated particle is small and thus the jet dynamics will be 
dominant. For this reason, any contribution from collective dynamics can 
be neglected in these results.

The modification observed in this analysis is found in a similar 
kinematic region to where the so-called ``Cronin'' peak has been 
observed. In the \xe bins where the broadened widths are observed, 
associated hadrons corresponding to trigger neutral pions in the range 
$5<\pttrig<9$ \gevc are approximately in the range $1<\ptassoc<2.5$ 
\gevc. The Cronin effect was once attributed to multiple scattering of 
partons within a nuclear medium; however, recent measurements revealed a 
particle ID dependence and have shown that additional final-state 
effects must also be present~\cite{Adare:2013esx}. Additional 
nonperturbative initial-state partonic \kt can also contribute to the 
Cronin peak, to which this measurement is sensitive. Nonetheless, 
multiple scattering interactions within the nucleus could manifest 
themselves as collisional energy loss or elastic scatterings leading to 
an angular broadening, both of which could lead to the observed 
away-side momentum width broadening in \pa collisions. Two-particle 
correlations may provide additional constraints on the underlying 
physical mechanism which leads to this phenomenon. Future measurements 
with particle identification will play an important role in identifying 
the cause of the Cronin peak, as a particle species dependence has been 
measured in $d$$+$Au~\cite{Adare:2013esx} collisions.

\begin{figure}[tbh]
	\includegraphics[width=1.0\linewidth]{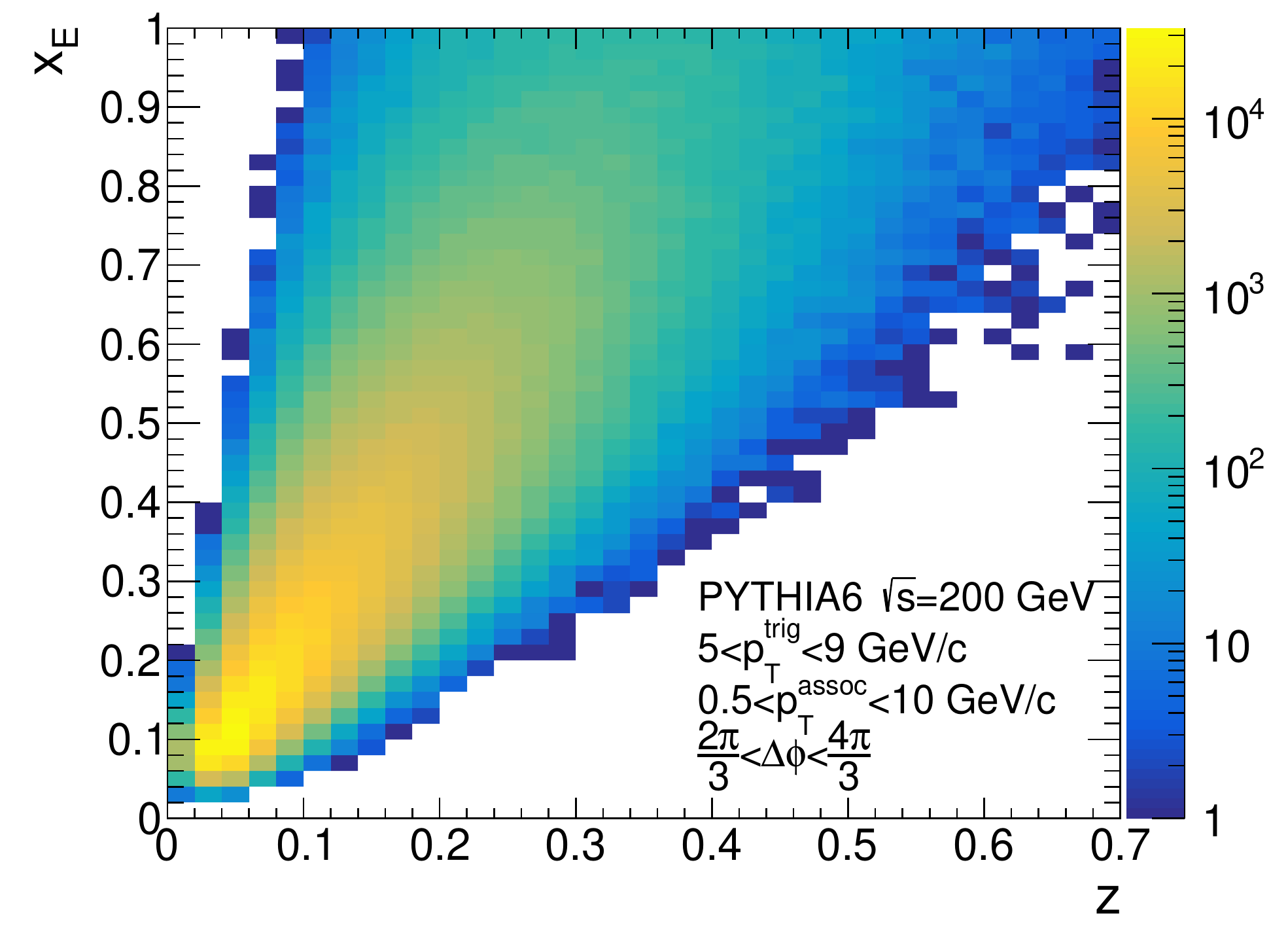}
\caption{The correlation between \xe and $z$ is shown as determined in a 
PYTHIA simulation for \pion-hadron correlations in the same kinematic 
regime as measured in the data. The correlation may provide insight into 
the origins of the inclusive hadron enhancement at moderate \pt in \pa 
collisions.}
    \label{fig:xevsz}
\end{figure}

In Ref.~\cite{Airapetian:2007vu} a strong dependence on $z$ to the 
inclusive charged hadron enhancement in $e$$+$$A$ collisions was found. In 
particular, the largest enhancement was found for $0.2<z<0.4$ hadrons. 
Figure~\ref{fig:xevsz} shows the correlation between \xe and $z$ for 
\pion-hadron correlations in \sqs=~200 GeV \pp collisions as determined 
in a PYTHIA6~\cite{Sjostrand:2006za} simulation with the default tune. 
The correlations are determined in the Monte Carlo simulation in the 
same kinematic regime as the data to draw a better comparison between 
$z$ and \xe to this analysis. Figure~\ref{fig:xevsz} shows that 
$0.25<\xe<0.5$, where the transverse momentum broadening is observed to 
be the largest, corresponds approximately to a range of $z$ covering 
$0.07<z<0.2$. This is in a similar region to where 
Ref.~\cite{Airapetian:2007vu} sees the largest inclusive hadron 
enhancement in $e$$+$$A$ collisions. However, it should also be considered 
that the two measurements cover a much different $Q^2$ range, which may 
also be relevant in this comparison.

Nuclear PDFs can also play a role; in particular the nuclear PDFs are 
known to vary with the longitudinal momentum fraction of the parton 
probed~\cite{Geesaman:1995yd,Hen:2016kwk}. These measurements are in a 
kinematic region that may be sensitive to the anti-shadowing region 
around $x\approx$0.1. The \pout correlation functions are also sensitive to 
a small transverse momentum scale, and thus can also probe the 
transverse-momentum-dependent parton distribution functions of the 
nucleus. Dijet measurements, where the jets have an average $\pt>4$ 
\gevc, have shown that there is a larger \dphi acoplanarity in \pa 
compared to \pp collisions~\cite{Corcoran:1990vq}, indicating that there 
is a nuclear size dependence to initial-state partonic transverse 
momentum in the kinematic regime where Cronin effects may be expected to 
be relevant. However, the observation from the present measurement that 
the broadening depends on $\Ncoll$ could indicate that the broadening is 
not simply due to additional transverse momentum from the nucleus size.

The transverse momentum broadening may also be due to soft radiative 
energy loss within the nucleus. Energy loss in cold nuclear matter has 
been previously studied with the Drell-Yan 
process~\cite{Vasilev:1999fa}. Transverse momentum broadening has also 
been measured to be nonzero in SIDIS 
interactions~\cite{Airapetian:2009jy}. While the Drell-Yan measurement 
is only sensitive to initial-state partonic energy loss, the SIDIS 
measurement and the measurement presented here are sensitive to both 
initial and final state energy loss. Global analyses which utilize all 
of these data may provide further insight into the origins of the 
measured transverse momentum broadening in nuclear Drell-Yan, SIDIS, and 
\pa$\rightarrow$~dihadrons processes. The difference in \pout between 
A+A and \pp collisions has been used to estimate the energy loss per 
unit length within the QGP in Au$+$Au 
collisions~\cite{Tannenbaum:2017afg}. Given that there is an observed 
difference in the \pout widths between \pa and \pp collisions this 
indicates that small energy losses have been measured in these dihadron 
correlations. Calculations for energy loss in a nucleus have been 
performed for both RHIC and LHC energies in the dijet and direct 
photon-hadron channel as well as in $e$$+$$A$ 
collisions~\cite{Xing:2012ii}.

\section{Conclusion}

In summary, high \pt dihadron correlations have been measured in 
\sqsntwo \pp, \pal, and \pau collisions. The \pout distributions are 
measured on the near and away side of the trigger hadron and the 
distributions are fit with Gaussian functions to extract the 
nonperturbative transverse momentum width in each system. The widths are 
compared across the various collision systems to search for 
modifications present in the nuclear collisions. No near-side 
modification is observed within uncertainties in the \pa collisions, 
indicating that intra-jet radiation effects from nuclei are small in 
these systems. In contrast, the away-side widths are broadened in \pau 
compared to \pp at moderate values of $x_E$, while no significant 
modification was observed in \pal compared to \pp. This was observed to 
be a function of the centrality or \Ncoll of the \pa collision, which 
suggests a path length dependence to the transverse momentum broadening.

A number of different physical effects may contribute to the measured 
transverse momentum broadening in \pa collisions. In particular, 
contributions from long range correlations were systematically studied 
and found to be small. The correlations are constructed in a kinematic 
regime where ``Cronin'' effects are known to be large; therefore, 
multiple initial-state scatterings or parton recombination effects in 
the final state may be contributing to the broadening. The correlations 
are also sensitive to the partonic initial-state transverse momentum, 
and thus may indicate additional primordial partonic $k_T$ in nuclei 
when compared to a free nucleon. However, the dependence of the 
broadening on \Ncoll suggests a path length dependence to hard scattered 
partonic energy loss, which may be due to radiative or elastic 
interactions with the nuclear remnants. Considering these different 
qualitative physics mechanisms, and the many different processes and/or 
observables with which they have been measured, will be an important 
endeavor in understanding hadronic interactions involving nuclei. Future 
measurements, especially at an electron-ion collider, will continue to 
shed light on the many physical phenomena that occur in proton-nucleus 
collisions.


\begin{acknowledgments}

We thank the staff of the Collider-Accelerator and Physics
Departments at Brookhaven National Laboratory and the staff of
the other PHENIX participating institutions for their vital
contributions.  We acknowledge support from the 
Office of Nuclear Physics in the
Office of Science of the Department of Energy,
the National Science Foundation, 
Abilene Christian University Research Council, 
Research Foundation of SUNY, and
Dean of the College of Arts and Sciences, Vanderbilt University 
(U.S.A),
Ministry of Education, Culture, Sports, Science, and Technology
and the Japan Society for the Promotion of Science (Japan),
Conselho Nacional de Desenvolvimento Cient\'{\i}fico e
Tecnol{\'o}gico and Funda\c c{\~a}o de Amparo {\`a} Pesquisa do
Estado de S{\~a}o Paulo (Brazil),
Natural Science Foundation of China (People's Republic of China),
Croatian Science Foundation and
Ministry of Science and Education (Croatia),
Ministry of Education, Youth and Sports (Czech Republic),
Centre National de la Recherche Scientifique, Commissariat
{\`a} l'{\'E}nergie Atomique, and Institut National de Physique
Nucl{\'e}aire et de Physique des Particules (France),
Bundesministerium f\"ur Bildung und Forschung, Deutscher Akademischer 
Austausch Dienst, and Alexander von Humboldt Stiftung (Germany),
J. Bolyai Research Scholarship, EFOP, the New National Excellence
Program ({\'U}NKP), NKFIH, and OTKA (Hungary),
Department of Atomic Energy and Department of Science and Technology 
(India),
Israel Science Foundation (Israel), 
Basic Science Research and SRC(CENuM) Programs through NRF
funded by the Ministry of Education and the Ministry of
Science and ICT (Korea).
Physics Department, Lahore University of Management Sciences (Pakistan),
Ministry of Education and Science, Russian Academy of Sciences,
Federal Agency of Atomic Energy (Russia),
VR and Wallenberg Foundation (Sweden), 
the U.S. Civilian Research and Development Foundation for the
Independent States of the Former Soviet Union, 
the Hungarian American Enterprise Scholarship Fund,
the US-Hungarian Fulbright Foundation,
and the US-Israel Binational Science Foundation.

\end{acknowledgments}



%
 
\end{document}